# Counting publications and citations:
# Is more always better?

Ludo Waltman, Nees Jan van Eck, and Paul Wouters

Centre for Science and Technology Studies, Leiden University, The Netherlands
{waltmanlr, ecknjpvan, p.f.wouters}@cwts.leidenuniv.nl

Is more always better? We address this question in the context of bibliometric indices that aim to assess the scientific impact of individual researchers by counting their number of highly cited publications. We propose a simple model in which the number of citations of a publication depends not only on the scientific impact of the publication but also on other 'random' factors. Our model indicates that more need not always be better. It turns out that the most influential researchers may have a systematically lower performance, in terms of highly cited publications, than some of their less influential colleagues. The model also suggests an improved way of counting highly cited publications.

## 1. Introduction

When bibliometrics is used for research assessment purposes, a general presumption seems to be that more is always better: The more publications, the better; the more citations, the better. At the same time, there is an increasing awareness that 'more is always better' should not be taken too literally. For instance, interpreting the number of citations of a publication as an *approximate* measure of the scientific impact of the publication, having more citations does not *always* coincide with having more impact. Publications with more citations may *on average* have more impact, but *individual* publications may deviate from this pattern. One could hypothesize, for instance, that authors of a publication tend to copy a substantial part of their reference list from the reference lists of earlier publications, often without paying serious attention to the contents of the referenced works (Simkin & Roychowdhury, 2003, 2005). If there is indeed some truth in this idea, it does not seem unlikely that publications sometimes become highly cited without actually having a lot of impact on subsequent scientific research. This illustrates that there does not exist a perfect relationship between scientific impact and citations. In addition to scientific impact, there are many other factors that may influence a publication's number of citations (Bornmann & Daniel, 2008; Martin & Irvine, 1983; Moed, 2005; Nicolaisen, 2007). Some of these factors are of a systematic nature, while others can be considered to have a more random character. In this paper, we are especially interested in these random factors.

Also when assessing the scientific impact of an oeuvre of publications rather than a single individual work, the more-is-better idea should be treated with care. It is not obvious, for instance, whether comparing the oeuvres of two researchers based on each researcher's total number of citations is a good approach. One researcher may have more citations than another researcher, but it could be that the latter researcher has authored a number of highly cited publications while the former researcher has earned his citations by producing an extensive oeuvre consisting exclusively of lowly and moderately cited works. In this situation, the researcher with the highly cited publications may actually have been more influential, despite his smaller overall number of citations. When assessing a researcher's scientific impact based on the total number of citations of his publications, the implicit assumption is that the number of citations of a publication is proportional to the scientific impact of the publication. This is a rather strong assumption. As argued by Ravallion and Wagstaff (2011), the true relationship between scientific impact and citations may well be non-linear.

In recent years, a large number of bibliometric indices were introduced that may serve as an alternative to counting a researcher's total number of citations. The best-known example is the *h*-index (Hirsch, 2005). This index is robust both to publications with only a small number of citations and to publications with a very large number of citations. This robustness is often considered a strong property of the *h*-index. Unfortunately, however, the *h*-index has other properties that are difficult to justify and that may cause inconsistencies in the results produced by the index (Waltman & Van Eck, 2012). An attractive alternative to the *h*-index is the highly cited publications (HCP) index (Bornmann, 2013; Waltman & Van Eck, 2012). This index counts the number of publications of a researcher that have received at least a certain minimum number of citations (e.g., Plomp, 1990, 1994). The HCP index has a similar robustness property as the *h*-index, but it does not suffer from the inconsistencies of this index.

In this paper, our focus is on the HCP index. The research question that we consider is whether more is always better when counting highly cited publications. To address this question, we introduce a simple model of the relationship between scientific impact and citations. The model shows that, as a consequence of random factors that influence the number of citations of a publication, the answer to our research question is negative. In itself, this may not be considered surprising. When working with small numbers of publications, it is to be expected that random factors may cause deviations from the more-is-better principle. For instance, a researcher with one highly cited publication need not always be more influential than a researcher who does not have any highly cited publications. However, our model reveals that random factors may result in deviations from the more-is-better principle that are of a systematic nature. These deviations occur even when dealing with large numbers of publications. In concrete terms, the model demonstrates how random effects may lead to paradoxical situations in which the most influential researchers have a systematically lower performance, in terms of highly cited publications, than some of their less influential colleagues. The model also suggests how the HCP index can be modified to avoid these paradoxical situations.

Before proceeding with our analysis, it is important to emphasize that the problem studied in this paper does not relate specifically to the HCP index. We focus on the HCP index because it is an important bibliometric index that, due to its simplicity, can be analyzed in a convenient way. However, findings similar to ours can be made for other bibliometric indices as well. Examples of such indices include the *h*-index (Hirsch, 2005) and its many variants, but also the generalizations of the HCP index recently proposed by Leydesdorff, Bornmann, Mutz, and Opthof (2011).

## 2. Scientific impact vs. citations

A crucial distinction in our analysis is between the scientific impact of a publication and the number of citations the publication has received. The scientific impact of a publication is the influence a publication has on subsequent scientific research. The number of citations of a publication partly reflects the scientific impact of the publication, but it also depends on a multitude of other factors (Bornmann & Daniel, 2008; Martin & Irvine, 1983; Moed, 2005; Nicolaisen, 2007). For instance, the number of citations of a publication may depend on the reputation of the authors, of the institutions with which the authors are affiliated, or even of the countries in which the authors are located. The citation behavior of researchers may play a role as well. If a researcher has a strong tendency to cite his own work, this obviously increases the number of citations of his publications. Scientific impact, reputation, and citation behavior are examples of factors that can be expected to have a systematic effect on the number of citations of a researcher's publications. If a researcher produces influential work, has a good reputation, or has a strong self citation tendency, this is likely to increase the number of citations of his publications in a systematic way.

The number of citations of a publication also depends on factors that can be considered to be more of a random nature (e.g., Dieks & Chang, 1976). Unlike the factors mentioned above, these random factors do not create a systematic advantage for the publications of one researcher compared with the publications of another research. It has been argued, for instance, that a substantial proportion of the references in a publication tend to be of a perfunctory nature (e.g., Moravcsik & Murugesan, 1975). These references are not essential for the citing publication but just serve to indicate that more work has been done on the same topic. The choice of perfunctory references tends to be quite arbitrary, since in many cases just a few publications are cited from a much larger set of publications that could all be cited equally well. Because of this arbitrariness, perfunctory references can be seen as a random factor influencing the number of times a publication is cited. Each researcher now and then benefits from perfunctory referencing, and there is no reason to expect the publications of one researcher to be advantaged in a systematic way over the publications of another researcher.

Although the choice of perfunctory references involves a significant degree of arbitrariness, one may expect that perfunctory references are more likely to refer to publications that already have a substantial number of citations than to publications with only a few citations. The former publications are more visible in the scientific literature and may therefore be more likely to receive additional citations. This would for instance be the case if researchers choose perfunctory references by more or less randomly selecting references from the reference lists of earlier publications (Simkin & Roychowdhury, 2003, 2005) or if researchers simply choose to refer to publications that are highly ranked by a search engine such as Google Scholar (i.e., a search engine that gives a substantial weight to citations to determine the ranking of publications). So random factors influencing the number of citations of a publication may create a self-reinforcing effect (often referred to as 'cumulative advantage', 'Matthew effect', or 'preferential attachment'; e.g., Price, 1976). The more citations a publication has, the more likely the publication is to receive additional citations.

## 3. More need not always be better

To address the question whether more is always better when counting highly cited publications, we introduce a simple model of the relationship between scientific impact and citations. Our model does not intend to provide an accurate representation of the many different factors influencing the number of citations of a publication. Instead, by introducing a number of simplifications, we aim to create an easy-to-understand model that still gives relevant insights into the more-is-better question.

In our model, we assume that scientific impact is the only systematic factor influencing the number of citations of a publication. Other systematic factors, such as reputation and citation behavior, are disregarded. Very importantly, however, we do incorporate in our model the idea that the number of citations of a publication may be influenced by random factors. To keep the model as simple as possible, we treat the scientific impact of a publication as a binary variable. A publication either does or does not have scientific impact. This is of course a highly unrealistic assumption. We will come back to this at the end of the paper.

We are interested in measuring researchers' overall scientific impact. We assume that the overall scientific impact of a researcher is determined by the number of high-impact publications the researcher has produced. We also assume that 10% of the publications in a scientific field have a high impact. The other 90% of the publications have a low impact. The scientific impact of low-impact publications is considered to be negligible.

The scientific impact of a publication cannot be directly observed, and we therefore look at the number of citations of a publication. We distinguish between two classes of publications: Publications that belong to the top 10% of their field in terms of citations and publications that, based on their number of citations, do not belong to the top 10% of their field. We refer to publications belonging to the top 10% most frequently cited of their field as highly cited publications.[1] Publications that do not belong to the top 10% most frequently cited of their field are referred to as lowly cited publications. Counting the number of highly cited publications of a researcher yields the above-mentioned HCP index.

---

[1] For the purpose of our analysis, practical difficulties in determining whether a publication belongs to the top 10% most frequently cited (Waltman & Schreiber, 2013) can be ignored.

In an ideal world in which there is a perfect relationship between the scientific impact of a publication and a publication's number of citations, being highly cited coincides with having a high impact. In other words, each highly cited publication is also a high-impact publication, and the other way around. In such an ideal world, the HCP index perfectly indicates the number of high-impact publications of a researcher, and the index therefore always provides a correct assessment of a researcher's overall scientific impact.

However, as we have discussed, the idea of a perfect relationship between scientific impact and citations is difficult to justify. In our model, random factors cause some publications to be highly cited even though they have only a limited scientific impact. Conversely, some publications do not belong to the top 10% most highly cited publications of their field even though they do belong to the 10% high-impact publications. A possible scenario is illustrated in Table 1. In this scenario, 3% of the publications in a field have a high impact and are also highly cited, while 7% of the publications have a high impact but are not highly cited and another 7% of the publications are highly cited but do not have a high impact. The remaining 83% of the publications have a low impact and are also lowly cited. In the scenario illustrated in Table 1, if a publication has a high impact, there is a probability of 3% / 10% = 0.30 that the publication is highly cited. If a publication has a low impact, this probability is just 7% / 90% ≈ 0.08. Hence, high-impact publications are (3% / 10%) / (7% / 90%) ≈ 3.86 times as likely to be highly cited as low-impact publications.

Table 1. Illustration of a scenario in which there is no perfect relationship between the scientific impact of a publication and a publication's number of citations.

|  | Lowly cited pub. | Highly cited pub. | Total |
| --- | --- | --- | --- |
| Low-impact pub. | 83% | 7% | 90% |
| High-impact pub. | 7% | 3% | 10% |
| Total | 90% | 10% | 100% |

In the scenario illustrated in Table 1, we may have the following interesting situation. Suppose we have two researchers, researcher A and researcher B (see Table 2). Researcher A has produced 100 publications, all of them of high impact. Researcher B has produced 500 publications, so five times as many as researcher A, but none of these publications is of a high impact.[2] Given our assumption that a researcher's overall scientific impact is determined by the number of high-impact publications the researcher has produced, we must conclude that researcher A has been highly influential while the scientific impact of researcher B has been negligible, despite the large publication output of this researcher.

Table 2. Four hypothetical researchers that are used to illustrate the consequences of different approaches to counting highly cited publications.

|  | Number of publications | | Number of publications | |
| --- | --- | --- | --- | --- |
|  | low-impact | high-impact | lowly cited | highly cited |
| Researcher A | 0 | 100 | 70 | 30 |
| Researcher B | 500 | 0 | 461 | 39 |
| Researcher C | 50 | 200 | 186 | 64 |
| Researcher D | 270 | 70 | 298 | 42 |

The interesting question is whether the HCP index confirms this conclusion. Given the percentages reported in Table 1, we can expect researcher A to have (3% / 10%) × 100 = 30 highly cited publications. For researcher B, the expected number of highly cited publications is (7% / 90%) × 500 ≈ 39. If researchers A and B indeed each have their statistically expected number of highly cited publications, we end up in the paradoxical situation in which the HCP index indicates that researcher B, with an HCP value of 39, appears to be more influential than researcher A, with an HCP value of 30. Hence, the HCP index provides an incorrect assessment of the overall scientific impact of the two researchers. Moreover, this incorrect assessment is not caused by an incidental statistical fluctuation. Since researchers A and B each have their statistically expected number of highly cited publications, the HCP index is systematically wrong in situations like ours.

Why does the HCP index in certain situations provide systematically incorrect assessments of researchers' overall scientific impact? This is because, as long as there is no perfect relationship between scientific impact and citations, a researcher with a given number of high-impact publications can always be outperformed, in terms of highly cited publications, by another researcher with a sufficiently large number of low-impact publications. Low-impact publications are less likely to become highly cited than high-impact publications, but by producing lots of low-impact publications it is still possible to obtain a large number of highly cited publications.

---

[2] In the theoretical examples presented in this paper, we know each publication's impact. This is helpful to illustrate our ideas. In practice, however, the impact of a publication cannot be directly observed.

The above scenario demonstrates that more need not always be better when counting highly cited publications. There can be systematic deviations from the more-is-better principle. In particular, the HCP index may overestimate the scientific impact of researchers who focus on producing lots of publications without paying much attention to the impact of their work.

Table 3. Scientific impact vs. citations. The parameter α determines the degree of correlation ($0 \leq \alpha \leq 0.09$).

|  | Lowly cited pub. | Highly cited pub. | Total |
|---|---|---|---|
| Low-impact pub. | $0.9 - \alpha$ | $\alpha$ | 0.9 |
| High-impact pub. | $\alpha$ | $0.1 - \alpha$ | 0.1 |
| Total | 0.9 | 0.1 | 1 |

Table 3 shows a generalization of the scenario illustrated in Table 1. The parameter α determines the degree to which scientific impact and citations are correlated. A perfect correlation is obtained by setting α equal to zero. The other extreme is to set α equal to 0.09, in which case scientific impact and citations are completely uncorrelated and the number of citations of a publication provides no indication at all of the scientific impact of the publication. The absence of any correlation between scientific impact and citations for α = 0.09 follows from the fact that setting α equal to 0.09 causes each cell in Table 3 to be equal to the product of the corresponding row and column totals, making scientific impact and citations statistically independent from each other. The possibility of setting α equal to a value above 0.09 can be ignored. This would lead to the implausible situation of a negative correlation between scientific impact and citations. Setting α equal to 0.07 yields the scenario illustrated in Table 1. In the end, the value of α that one considers most realistic depends on how much trust one has in the ability of citations to indicate the scientific impact of a publication. It also depends on the exact interpretation that one gives to the notion of scientific impact. Moreover, since citation cultures differ across scientific fields, it may well be that different fields require different values of $\alpha$.

Based on Table 3, it can be seen that producing $n_{HI}$ high-impact publications on average yields $[(0.1 - \alpha) / 0.1] \times n_{HI}$ highly cited publications. Similarly, producing $n_{LI}$ low-impact publications on average yields $[\alpha / 0.9] \times n_{LI}$ highly cited publications. It follows that obtaining a single highly cited publication on average requires $1 / [(0.1 - \alpha) / 0.1]$ high-impact publications or $1 / [\alpha / 0.9]$ low-impact publications. Clearly, the lower the value of $\alpha$, the more the HCP index rewards the production of high-impact publications. Nevertheless, for any non-zero value of $\alpha$, a researcher with a given number of high-impact publications can be systematically outperformed, in terms of highly cited publications, by a researcher with lots of low-impact publications. More precisely, a researcher who produces more than $[(0.1 - \alpha) / 0.1] / [\alpha / 0.9] \times n_{HI} = (0.9 - 9\alpha) / \alpha \times n_{HI}$ low-impact publications on average outperforms a colleague producing $n_{HI}$ high-impact publications. Of course, if the value of $\alpha$ is close to zero, the number of low-impact publications required to outperform a researcher with $n_{HI}$ high-impact publications becomes very large, and in practice it may not be possible to have such a large publication output.

## 4. An improved counting approach

An obvious question is whether the HCP index can be modified in such a way that it no longer suffers from systematic errors in the assessment of researchers' overall scientific impact. In other words, is it possible to develop an improved way of counting highly cited publications?

One possibility might be to move from a size-dependent HCP index to a size-independent one. In that case, instead of calculating the *number* of highly cited publications of a researcher, one would calculate a researcher's *proportion* of highly cited publications. In some situations, this would indeed lead to improved results. For instance, consider the scenario illustrated in Table 1, and take the situation of researchers A and B, as discussed in the previous section (see Table 2). Researcher A has produced 100 high-impact publications, of which 30 are highly cited. Researcher B has produced 500 low-impact publications, of which 39 are highly cited. As we have seen, when looking at a researcher's number of highly cited publications, researcher B outperforms researcher A, even though researcher B's scientific impact is negligible compared with researcher A's. Now suppose we look at the proportion of highly cited publications of a researcher, that is, a researcher's number of highly cited publications divided by his total number of publications. Researcher A has 30 / 100 = 30% highly cited publications, while researcher B has only 39 / 500 = 7.8% highly cited publications. Hence, when looking at a researcher's proportion of highly cited publications, researchers A and B are ranked correctly with respect to each other.

Unfortunately, a size-independent HCP index also has problems. To demonstrate this, we introduce a third researcher, researcher C. Suppose researcher C has produced 200 high-impact publications and 50 low-impact ones (see Table 2). In line with the percentages reported in Table 1, this has resulted in (3% / 10%) × 200 + (7% / 90%) × 50 ≈ 64 highly cited publications. Since researcher C has produced twice as many high-impact publications as researcher A, researcher C's scientific impact is also twice as large as researcher A's. However, researcher A has 30% highly cited publications, while researcher C has only 64 / (200 + 50) = 25.6% highly cited publications. Hence, according to a size-independent HCP index, researcher A outperforms researcher C. It is clear that this is an incorrect assessment of the scientific impact of the two researchers.

From the point of view of assessing researchers' overall scientific impact, the fundamental problem of a size-independent HCP index is that productivity is not rewarded. If two researchers have the same proportion of highly cited publications, their scientific impact is assessed to be the same as well. This makes no sense if one researcher for instance has a publication output twice as large as another researcher. Other things being equal, the overall scientific impact of a researcher should be assessed proportionally to his publication output.[3] If one researcher has both twice as many highly cited and twice as many lowly cited publications as another researcher, then the scientific impact of the former researcher should be assessed to be twice as large as the scientific impact of the latter researcher. A size-independent HCP index fails to take such productivity considerations into account.

---

[3] In practice, other things need not always be equal. For instance, one researcher may have more research time than another. For the purpose of our analysis, however, we assume researchers to find themselves in comparable situations.

There turns out to be a better way in which the HCP index can be modified to make sure that it provides proper assessments of researchers' scientific impact. The HCP index can be seen as a weighted sum of the publications of a researcher, where a highly cited publication has a weight of one while a lowly cited publication has a weight of zero. We now show that the weights used in the HCP index can be modified in such a way that on average the HCP value of a researcher is exactly equal to the number of high-impact publications the researcher has produced.

Our starting point is the general scenario shown in Table 3, with the parameter $\alpha$ ($0 \leq \alpha \leq 0.09$) determining the degree to which scientific impact and citations are correlated. We propose to weight highly cited publications by

$$w_{\text{HC}} = \frac{0.1\alpha - 0.09}{\alpha - 0.09} \tag{1}$$

and lowly cited publications by

$$w_{\text{LC}} = \frac{0.1\alpha}{\alpha - 0.09}. \tag{2}$$

Hence, the HCP value of a researcher is given by

$$\text{HCP} = n_{\text{LC}} w_{\text{LC}} + n_{\text{HC}} w_{\text{HC}}, \tag{3}$$

where $n_{\text{LC}}$ and $n_{\text{HC}}$ denote the number of lowly and highly cited publications of the researcher. Notice that setting $\alpha$ equal to zero yields $w_{\text{HC}} = 1$ and $w_{\text{LC}} = 0$, which means that (3) reduces to the standard HCP index discussed in the previous section. Notice also that $w_{\text{HC}}$ and $w_{\text{LC}}$ are not defined if $\alpha$ is set equal to 0.09. As we have seen in the previous section, if $\alpha$ is set equal to 0.09, the number of citations of a publication does not provide any indication of the scientific impact of the publication.

Suppose a researcher has produced $n_{\text{HI}}$ high-impact publications and $n_{\text{LI}}$ low-impact publications. The expected HCP value of the researcher calculated using (1), (2), and (3) then equals $n_{\text{HI}}$. This can be seen as follows. Based on Table 3, we obtain

$$\mathrm{E}(n_{\mathrm{HC}}) = \frac{0.1-\alpha}{0.1} n_{\mathrm{HI}} + \frac{\alpha}{0.9} n_{\mathrm{LI}} \tag{4}$$

and

$$\mathrm{E}(n_{\mathrm{LC}}) = \frac{\alpha}{0.1} n_{\mathrm{HI}} + \frac{0.9-\alpha}{0.9} n_{\mathrm{LI}}, \tag{5}$$

where E(•) denotes the expected value operator. It follows from (3) that

$$\mathrm{E}(\mathrm{HCP}) = \mathrm{E}(n_{\mathrm{LC}}) w_{\mathrm{LC}} + \mathrm{E}(n_{\mathrm{HC}}) w_{\mathrm{HC}}. \tag{6}$$

Substitution of (1), (2), (4), and (5) into (6) results in

$$\mathrm{E}(\mathrm{HCP}) = n_{\mathrm{HI}}. \tag{7}$$

This proves that on average the HCP value of a researcher calculated using (1), (2), and (3) is exactly equal to the number of high-impact publications the researcher has produced. Unlike the standard HCP index, our modified HCP index therefore does not suffer from systematic errors in the assessment of researchers' scientific impact.

To understand the mechanism of our modified HCP index, it is important to see that $w_{\mathrm{LC}}$ in (2) is always negative (except if $\alpha$ is set equal to zero). Hence, lowly cited publications are given a negative weight in our modified HCP index. Other things equal, the more lowly cited publications one has, the lower one's HCP value. Why do we give a negative weight to lowly cited publications? Given our assumption that the scientific impact of low-impact publications is negligible, we want the contribution of a low-impact publication to a researcher's HCP value to be zero on average. However, due to random factors influencing the number of citations of a publication, some low-impact publications end up being highly cited, and these publications make a positive contribution to a researcher's HCP value. To compensate for this, we give a negative weight to lowly cited publications. This negative weight is chosen in such a way that on average the contribution of a low-impact publication to a researcher's HCP value is zero. For a high-impact publication, we want the contribution to a researcher's HCP value to be one on average. Using the weights in (1) and (2), we accomplish both of our objectives: Low-impact publications make an average contribution of zero, and high-impact publications on average contribute one.

Finally, there is an interesting property of our modified HCP index that we want to demonstrate. We again consider the scenario illustrated in Table 1. Let us introduce a new researcher, researcher D. Suppose this researcher has produced 70 high-impact publications and 270 low-impact ones (see Table 2). In this way, he has obtained the expected number of (3% / 10%) × 70 + (7% / 90%) × 270 = 42 highly cited publications. His remaining 70 + 270 – 42 = 298 publications are lowly cited. Setting α equal to 0.07 in (1) and (2), we obtain $w_{HC}$ = 4.15 and $w_{LC}$ = –0.35. Using (3), we then find that the HCP value of researcher D equals 298 × (–0.35) + 42 × 4.15 = 70. Hence, as expected, researcher D's HCP value equals his number of high-impact publications. A similar calculation can be made for researcher A introduced earlier (see Table 2). Recall that this researcher has produced 100 high-impact publications, which has resulted in 30 highly cited publications and 70 lowly cited ones. Based on his number of highly and lowly cited publications, we obtain a HCP value of 100 for researcher A, which is exactly the number of high-impact publications this researcher has produced. Comparing researchers A and D, our modified HCP index correctly identifies researcher A as the one with the larger scientific impact.

What is interesting in the comparison of researchers A and D is that researcher A is outperformed by researcher D in terms of both highly cited publications (30 vs. 42) and lowly cited publications (70 vs. 298). Intuitively, this may seem sufficient evidence to conclude that researcher D must have a larger scientific impact than researcher A. However, as we have seen, researcher A is the one with the larger scientific impact. Hence, based on simple more-is-better logic, one would easily draw an incorrect conclusion in the comparison of researchers A and D. By deviating from the more-is-better logic, our modified HCP index reaches the correct conclusion.

## 5. Discussion and conclusion

The more-is-better principle plays a central role in evaluative bibliometrics. In this paper, we have given examples of situations in which more need not always be better. When the overall scientific impact of researchers is determined by their number of high-impact publications, having more highly cited publications need not always coincide with having a larger scientific impact. This is caused by random factors that may influence the number of citations of a publication. The stronger these random factors, the more difficult it becomes to maintain the more-is-better principle. Importantly, the deviations from the more-is-better principle that we have studied are of a systematic nature. They do not simply result from incidental statistical fluctuations. This shows that, contrary to what sometimes seems to be claimed (e.g., Van Raan, 1998), random effects on citations need not cancel out. Instead, random effects may have systematic consequences, at least when using certain types of bibliometric indices.

The model that we have analyzed in this paper is extremely stylized. On the one hand this makes the model easy to study, but on the other hand it also means that the model has significant weaknesses. The most important weakness may be that the scientific impact of a publication is assumed to be a binary variable: A publication either does or does not have scientific impact. Although this is of course a highly unrealistic assumption, it does match well with the idea of counting highly cited publications, which also relies on a binary distinction, albeit based on citations rather than impact.[4] Future work could focus on constructing more detailed models of the relationship between scientific impact and citations to find out under what types of conditions our findings do or do not remain valid.

We emphasize that we consider the modified HCP index introduced in Section 4 to be mainly of theoretical interest. To obtain appropriate weights for lowly and highly cited publications, one would need to have a realistic value for the parameter α. It is not evident how such a value could be determined empirically. Moreover, our modified HCP index is completely based on our very simple model of the relationship between scientific impact and citations. This makes the index vulnerable to the weaknesses of this model.

Nevertheless, we do believe that our modified HCP index provides interesting insights. The index illustrates how random effects on the number of citations of a publication can be corrected for while staying within the framework of simple additive indices with their many attractive properties (Marchant, 2009; Ravallion &

---

[4] By assuming a binary concept of scientific impact, our model serves as a kind of ideal world for the HCP index. In a model with a continuous concept of impact, it would be fundamentally impossible for the HCP index to provide perfect measurements of impact. In a model with a binary concept of impact, it is theoretically possible for the HCP index to provide perfect measurements of impact, as we have shown in Section 4.

Wagstaff, 2011). In addition, our modified HCP index introduces the idea of giving a negative weight to certain publications, not because these publications have a 'negative impact', but simply as a kind of correction factor to ensure that the index on average produces correct results. We emphasize that the insights we have obtained for HCP indices may be applicable to other bibliometric indices as well.

We hope that this paper will stimulate more research into the development of bibliometric indices within a model-based framework, in particular within a framework in which the relationship between citations on the one hand and concepts such as scientific impact and scientific quality on the other hand is made explicit (see also Ravallion & Wagstaff, 2011).

## Acknowledgments

We would like to thank two anonymous referees for their comments on an earlier version of this paper. We are also grateful to our colleagues at the Centre for Science and Technology Studies for their feedback on our work.